\documentclass[twocolumn,prd,reprint,amsmath,amssymb,superscriptaddress,floatfix]{revtex4}

\pdfoutput=1
\usepackage[utf8]{inputenc}
\usepackage{amssymb}
\usepackage{amsmath}
\usepackage{amsfonts}
\usepackage{graphicx}
\usepackage{color}
\usepackage{xspace}
\usepackage{comment}
\usepackage{hyperref}
\usepackage[normalem]{ulem}
\usepackage[section]{placeins}
\usepackage{afterpage}
\usepackage{slashed}
\usepackage{enumerate}
\usepackage{enumitem}
\usepackage{caption}
\usepackage{subfigure}
\usepackage{float}
\usepackage{ulem}
\usepackage{verbatim}
\usepackage{multirow,rotating}
\usepackage[dvipsnames]{xcolor}
\usepackage{braket}

\definecolor{dcolour}{rgb}{.5, .5, .5}
\def\gsim{\raise0.3ex\hbox{$\;>$\kern-0.75em\raise-1.1ex\hbox{$\sim\;$}}}
\def\lsim{\raise0.3ex\hbox{$\;<$\kern-0.75em\raise-1.1ex\hbox{$\sim\;$}}}
\def\gsim{\raise0.3ex\hbox{$\;>$\kern-0.75em\raise-1.1ex\hbox{$\sim\;$}}}
\def\lsim{\raise0.3ex\hbox{$\;<$\kern-0.75em\raise-1.1ex\hbox{$\sim\;$}}}

\newcommand{\ba}[1]{\begin{eqnarray} \label{(#1)}}
	\newcommand{\ea}{\end{eqnarray}}

\newcommand{\red}[1]{\textcolor{red}{#1}}
\usepackage{ulem} 

\begin{document}

\title{Exploring the molecular scenario of $X(4160)$}

\author{Ruitian Li}
\email{liruitian@mail.dlut.edu.cn}
\author{Dazhuang He}
\email{dzhe@mail.dlut.edu.cn}
\affiliation{Institute of Theoretical Physics, School of Physics, Dalian University of Technology, \\ 
No.2 Linggong Road, Dalian, Liaoning, 116024, People’s Republic of China}
\author{Xuan Luo}
\email{xuanluo@ahu.edu.cn}
\affiliation{School of Physics and Optoelectronics Engineering, Anhui University, \\
Hefei, Anhui 230601, People’s Republic of China}
\author{Hao Sun}
\email{haosun@dlut.edu.cn}
\affiliation{Institute of Theoretical Physics, School of Physics, Dalian University of Technology, \\ 
No.2 Linggong Road, Dalian, Liaoning, 116024, People’s Republic of China}


\begin{abstract}

In this work, we investigate the decays of meson $X(4160)$ in the molecular frame. 
By using an effective lagrangian approach, we calculate the widths of $X(4160)\to J/\psi\phi$, $D^*\bar{D}^*$, $D^*\bar{D}$ and $D\bar{D}$ 
through the triangle loop mechanism in the molecular scenario. Based on the compositeness condition, the coupling between $X(4160)$ and the $D_s^*\bar{D}_s^*$ molecular state is determined.
The other coupling constants used are determined phenomenologically. 
The calculation of the decay widths shows that the $D^*\bar{D}^*$ decay mode has a larger branching ratio. 
These predictions can be seen as a test for the charmonium assignment of $X(4160)$.

\end{abstract}
\vskip10mm
\maketitle
\flushbottom
%
%

\section{Introduction}
\label{I}

Hadron physics plays an important role in physics, focusing on the exploration of strong interactions. 
In the past decade, many resonance states 
that cannot be explained by the conventional quark model~\cite{ParticleDataGroup:2014cgo},
have been experimentally discovered.
In the framework of quantum chromodynamics, it is allowed to exist some exotic states beyond the quark model,  
such as multi-quark states~\cite{Maiani:2014aja,Maiani:2013nmn,Esposito:2016noz,Fleming:2007rp,Mehen:2011ds,Wang:2013kra,Gamermann:2009fv,Wu:2021ezz}, the hybrid state~\cite{Chanowitz:1982qj,BARNES1983241},  glueballs~\cite{Klempt:2007cp,Esposito:2016noz} and so on.
Further study of these exotic states can provide us with a great help in revealing the nature of strong interactions. Nowadays, a large number of $XYZ$ states have been observed experimentally, which are called charmonium-like states, 
as the common feature is that their decay final states contain $c$ and $\bar{c}$ quarks~\cite{Brambilla:2019esw,Cleven:2015era,Ma:2014zva,Liu:2019zoy,Hosaka:2016pey}. 
Most of the $XYZ$ states appear at certain hadronic thresholds, 
a wide range of hadronic molecule explanations for these states have been presented~\cite{Guo:2017jvc}.

The first exotic state, $X(3872)$, was discovered by the Belle Collaboration 
in the $J/\psi \pi^+ \pi^-$ invariant mass spectrum of the $B^+ \to K^+ + J/\psi\pi^+\pi^-$ process in 2003~ \cite{Belle:2003nnu},
which was further confirmed by the BaBar Collaboration~\cite{BaBar:2004oro} in the same decay channel. 
The mass of $X(3872)$ is near the threshold of $D^0 \bar{D}^{*0}$, 
which makes it a natural candidate for the $D^0 \bar{D}^{*0}$ molecule-like state.
The same assumption works for the $X(4160)$ particle.
The particle $X(4160)$ was first observed from the process $e^+e^{-} \to J/\psi D^* \bar{D}^*$ in the $D^*\bar{D}^*$ mass spectrum by the Belle Collaboration~\cite{Belle:2007woe}, and then by the LHCb Collaboration in the $J/\psi\phi$ mass spectrum from the process of $B^+ \to J/ \psi \phi K^+$~\cite{LHCb:2021uow}.
The mass and width of $X(4160)$ are $M = (4153^{+23}_{-21}) $ MeV and $\Gamma = (136^{+60}_{-61} )$ MeV, respectively.
It's quantum numbers are not yet well know.

With the experimental discoveries, a lot of works
have been and are being research on the $X(4160)$ resonance.
In the framework of QCD sum rules, they investigated the mass spectrum of hidden charm tetraquark states and suggests that $X(4160)$ could be eitherthe scalar or tensor $qc\bar{q}\bar{c}$ tetraquark state~\cite{Chen:2017dpy}.
In Ref.~\cite{Yang:2009fj}, the authors studied the properties and strong decay width of $X(4160)$ as $\chi_{c0}(3^3P_0)$, $\chi_{c1}(3^3P_0)$, $\eta_{c2}(2^1D_2)$ or $\eta_{c2}(2^1D_2)$ candidate by the 3P0 model, respectively.
However, $X(4160)$ as $\eta_{c}(4S)$ candidate is in conflict with the prediction of the mass and decay width of $\eta_{c}(4S)$~\cite{He:2014xna}.
In Ref.~\cite{Li:2009zu}, by studing the influence of screened potential on the spectrum of charmoniums, 
the authors believed that $X(4160)$ was a more appropriate state for $\chi_{c0}(3P)$. 
In Ref.~\cite{Wang:2022dfd}, by using Bethe-Salpeter method, 
the authors calculated the charmonium state decays of $2P$ and $3P$. 
By solving the Salpeter equation, they obtained $M_{\chi_{c0}(3P)} = 4.140$ GeV and $\Gamma_{\chi_{c0}(3P)} = 81$ MeV,
which are very close to the mass and decay width of $X(4160)$. The result supports X(4160) is a candidate for the $\chi_{c0}(3P)$.

Since the measured mass of $X(4160)$ is close to the threshold of $D^*_s\bar{D}^*_s$, hadronic molecule explanation can also make sense. 
It is worth mentioning that before 2017 many researchers tried to identify $X(4140)$ as a $D_s^*D_s^*$ molecule with possible quantum numbers of $0^{++}$ and $2^{++}$~\cite{Liu:2009ei,Branz:2009yt,Chen:2015fdn,Karliner:2016ith}. It is considered that since the light meson channel is not considered in their work making the width of the state small, so the association with the experimentally obtained smaller width of $X(4140)$ is more natural. Finally, with the quantum number of $X(4140)$ was determined to be $1^{++}$~\cite{ParticleDataGroup:2018ovx}, the $D^*_sD_s^*$ molecule is more related to $X(4160)$.

By using hidden gauge theory, the authors obtained the $D_s^*\bar{D}^*_s$ molecular state with $J^{PC}=2^{++}$ associated with $X(4160)$~\cite{Molina:2009ct}. Then in Ref.~\cite{Wang:2017mrt}, the authors researched the $J/\psi\phi$ mass distribution of the $B^+\to J/\psi \phi K^+$ reaction, considering the $D_s^*\bar{D}^*_s$ molecular state $X(4160)$ with $J^{PC}=2^{++}$ based on the Ref.~\cite{Molina:2009ct} and resonance $X(4140)$, which could well explain the experimental data. 
There were also works from QCD sum rules studied that favored the $D_s^*\bar{D}^*_s$ structure with quantum number $2^{++}$ for the $X(4160)$ state~\cite{MartinezTorres:2016cqv,Albuquerque:2018jkn}. In many other works~\cite{Ikeno:2018ugx,Wang:2018djr}, the $X(4160)$ state, as a $D_s^*\bar{D}^*_s$ molecular state, used the quantum numbers of $J^{PC}=2^{++}$. 
There are some previous works focusing on the study of $D_s^*\bar{D}_s^*$ molecular state~\cite{Dong:2020hxe,Ding:2021igr,Ji:2022vdj}.

Until now,
the nature of meson $X(4160)$ has not yet been identified.
Stimulated by this discovery,
we study the decay of $X(4160)$ as a molecular state to two vector meson final states 
and calculate the decay widths of two decay channels by using the effective lagrangian approach. 
In our study, we treat $X(4160)$ as a molecular state with quantuam number $2^{++}$, which is based on the prediction in Ref.~\cite{Molina:2009ct}.
The channel of $X(4160)$ decays to $J/\psi\phi$ is produced by exchanging the mesons of $D_{s0}$, $D_{s}$ and $D_s^*$. 
And the channel of $X(4160) \to D^*\bar{D}^*/D^*\bar{D}/D\bar{D}$ is considered by exchanging the mesons of $K$ and $K^*$. 
In addition to research mass spectrum, the study of the decay behavior is also essential and presented to understand the properties of $X(4160)$. 

The present paper is assigned as follows. 
The effective lagrangians and the decay formula are given in Sec.~\ref{sec2}. 
Sec.~\ref{III} shows our numerical results and discussion. 
The summary is presented in the last section.

\section{framework}	
\label{sec2}

\subsection{Molecular structure}
\label{A}

\begin{figure}[h]
	\begin{center}
		\includegraphics[scale=0.5]{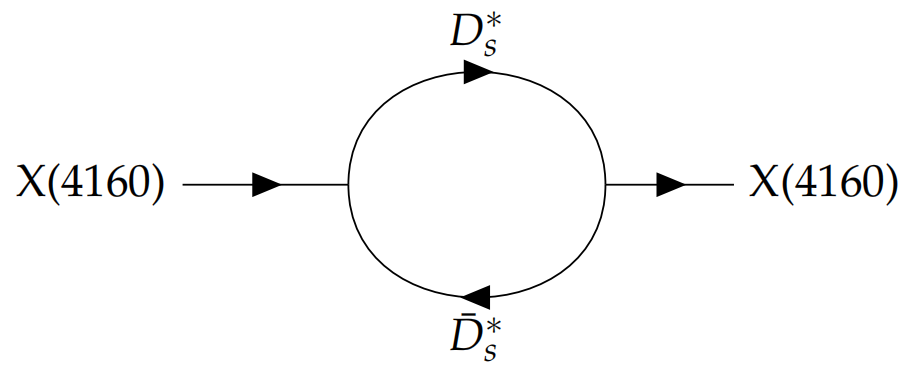}
	\end{center}
	\vspace{-0.7cm}
	\caption{The mass operator of the $X(4160)$ }
	\label{fig1}
\end{figure}

According to the effective Lagrangian method, the decay width of the strong decay process involving the coupling of $D_s^*\bar{D}^*_s$ molecule state with $X(4160)$ hadron is determined.
  We give the following effective lagrangian to describe the interaction between hadron state and their components \cite{Xiao:2020ltm},
\begin{equation}\label{eq1}
\begin{aligned}
\mathcal{L}_{X}(x)= g_{X} X^{\mu \nu}(x) \int dy \Phi(y^2) \bar{D}_{s}^{*\mu}(x+\omega_2y) {D_{s}^{*}}^{\nu}(x-\omega_1 y),
\end{aligned}
\end{equation}
where $y$ is the relative Jacobi coordinate and the tensor field is represented by $X^{\mu \nu}(x)$. $\omega_1$ and $\omega_2$ are mass ratios as defined below, 
\begin{equation}
\omega_1=\frac{m_{\bar{D}^{*}_s}}{m_{D^*_s}+m_{\bar{D}_s^*}} \qquad \omega_2=\frac{m_{D^*_s}}{m_{D^*_s}+m_{\bar{D}^*_s}}.
\end{equation}	
From the expressions we can clearly see that the values of $\omega_1$ and $\omega_2$ are equal and their values both are $1/2$. $g_{X}$ is a coupling constant between $D^*_s \bar{D}^*_s$ molecule and the corresponding state $X(4160)$, its dimension is GeV. In order to describe the distribution of components in the molecular picture, the correlation function $\Phi\left(y^2\right)$ needs to be introduced. Since the self-energy function of Feynman diagram has ultraviolet (UV) divergence, we need to choose an appropriate form for the correlation function to offset the dispersion. The Fourier transform of the correlation function form coordinate space to momentum space takes the following form:
\begin{equation}
\Phi (y^{2})=\int \frac{d^{4} p}{(2 \pi)^{4}} e^{-i p y} \tilde{\Phi}(-p^{2}).
\end{equation}
We choose the Gaussian form of the correlation function, which can make the UV limited in the UV region of Euclidean space \cite{Dong:2014zka,Gutsche:2014zda,Faessler:2007gv,Dong:2013iqa}:
\begin{equation}\label{eq2}
\tilde{\Phi}(p_E^2)=\exp \left({-\frac{p_{E}^2}{\Lambda^2}}\right),
\end{equation}
where $\Lambda$ is a free size parameter and $P_{E}$ is the Jacobi momentum in Euclidean space. This vertex function is also widely used to estimate the decays of hadronic molecule~\cite{Weinberg:1962hj,Faessler:2007gv,Faessler:2007us,Branz:2009yt}. Following Refs.~\cite{Weinberg:1962hj,Efimov:1993zg,Faessler:2007gv
}, when $\Lambda$ is ensured, the coupling $g_X$ can be determined by the Weinberg-Salam compositeness condition \cite{Weinberg:1962hj,Salam:1962ap}.
The compositeness condition is,
\begin{equation}\label{eq5}
	\begin{aligned}
		Z=&1- \frac{\partial \Sigma_X^{(1)}(p^2)}{\partial p^2} |_{{p^2}=M_{(X)}^2} =0 .
	\end{aligned}
\end{equation}
where $\Sigma_{{X}}^{(1)}(p^2)$ is the remainder of the tensor $X(4160)$ particle mass operator $	\Sigma^{\mu\nu,\alpha\beta}\left(p^2\right)$.
It suggests that the renormalization constant of the meson as hadronic molecule state wave function is set to zero. In FIG.~\ref{fig1}, we shows the Feynman diagram contributing to the mass operator of $X(4160)$, where $X(4160)$ is regarded as the hadronic molecular state of $D_s^*\bar{D}^*_s$.

For spin-parity $2^+$ tensor meson $X\left(4160\right)$, $\Sigma^{\mu\nu,\alpha\beta}_{(p^2)}$ is defined as
\begin{equation}
\begin{aligned}
	&\Sigma^{\mu\nu,\alpha\beta}_{(p^2)} \\
	=& \frac{1}{2}\left(g_{\mu\alpha}g_{\nu\beta}+g_{\mu\beta}g_{\nu\alpha}\right)\Sigma^{(1)}_{X}\left(p^2\right) \\
	+&g_{\mu \nu} g_{\alpha \beta} \Sigma^{(2)}_{X}\left(p^2\right) \\
	+&\frac{1}{2}\left[g_{\mu \nu} p_{\alpha} p_{\beta}+g_{\alpha \beta} p_{\mu} p_{v}\right]\Sigma^{(3)}_{X}\left(p^2\right) \\
	+&\frac{1}{4}\left[g_{\mu \alpha} p_{v} p_{\beta}+g_{\mu \beta} p_{v} p_{\alpha}+g_{v \alpha} p_{\mu} p_{\beta}+g_{v \beta} p_{\mu} p_{\alpha}\right] \Sigma^{(4)}_{X}\left(p^2\right)\\
	+& p_{\mu} p_{v} p_{\alpha} p_{\beta} \Sigma^{(5)}_{X}\left(p^2\right),
\end{aligned}
\end{equation}
where only $\Sigma^{(1)}_{X}\left(p^2\right)$ is contribute to the mass renormalization of $X\left(4160\right)$.

By applying the following Lorentz structure to the $\Sigma^{\mu\nu,\alpha\beta}(p^2)$ we can directly obtain the scalar term $\Sigma^{(1)}_{(p^2)}$ that is useful for the wave function \cite{Xiao:2020alj}:
\begin{equation}
T_{\perp}^{\mu v \alpha \beta}=\frac{1}{10}\left(P_{\perp}^{\mu \alpha} P_{\perp}^{v \beta}+P_{\perp}^{\mu \beta} P_{\perp}^{v \alpha}\right)-\frac{1}{15} P_{\perp}^{\mu v} P_{\perp}^{\alpha \beta},
\end{equation}
the $P^{\mu\nu}_{\perp}$ is defined as $P_{\perp}^{\mu v}=g^{\mu v}-p^{\mu} p^{v} / p^{2}$. Eventually we use the following equation to obtain the scalar function $\Sigma^{\left(1\right)}\left(p^2\right)$:
\begin{equation}
\Sigma^{(1)}(p^{2})=T_{\perp}^{\mu v \alpha \beta} \Sigma^{\mu\nu,\alpha\beta}_{(p^2)}.
\end{equation}
By using the Lagrangian, we get the expression for the mass operator in FIG.~\ref{fig1} and present below: 
\begin{equation}
	\begin{aligned}
		\Sigma^{\mu\nu,\alpha\beta}_{X(4160)}&=g_{X}^2\int \frac{d^4k}{(2\pi)^4} \tilde{\Phi}^2(-(k-\frac{p}{2}))\\
		&\times \frac{-g^{\mu \nu }+\frac{k^{\mu}k^{\nu}}{m_{D_s^*}^2}}{m_{D^*_s}^2-k^2}\frac{-g^{\alpha\beta}+\frac{(k^{\alpha}-p^{\alpha})(k^{\beta}-p^{\beta})}{m_{\bar{D}^*_s}^2}}{m_{\bar{D}_s^*}^2-(k-p)^2},
	\end{aligned}
	\label{eq7}
\end{equation}
where $p^2=m_X^2$ , $p$ and $k$ are four momentums and $m_X$ and $m_{D_s^*}$ are the mass of $X(4160)$ and $D^*_s$ respectively.

We constructed the effective lagrangians describing the interaction between the $X(4160)$ molecular state with quantum number $1^+$  and its components,
\begin{equation}\label{eq10}
	\begin{aligned}
		\mathcal{L}_{X_1}(x)&= ig_{X_1} \epsilon_{\mu \nu \alpha \beta} \partial^{\mu} X_1^{\nu}(x) \int dy \Phi(y^2) \bar{D}_{s}^{*\alpha}(x+\omega_2y)\\ &{D_{s}^{*}}^{\beta}(x-\omega_1 y),
	\end{aligned}
\end{equation}
where $X_1$  refer to $X(4160)$ with $J^P$ of$1^+$, the others are the same as the former definitions.

\subsection{Decays of $X(4160) \to J/\psi \phi$}
\label{B}

In this part, we intend to be calculating the two body decays of $X(4160) \to J/\psi\phi$ by exchanging the mesons of $D_{s0}$ and $D_s$. Besides the Lagrangian given in Eq.~(\ref{eq1}), we need to use the others, which are listed below,
\begin{equation}\label{eq14}
	\mathcal{L}_{D_{s}^{*} \bar{D}_{s 0} V}=g_{V \bar{D}_{s 0} D_{s}^{*}}\left[D_{s 0}^{-} D_{s ; \mu v}^{*+}-D_{s 0}^{+} D_{s ; \mu v}^{*-}\right] V^{\mu v},
\end{equation}
where $D^*_{s;\mu\nu}=\partial_{\mu}D^*_{s;\nu}-\partial_{\nu}D^*_{s;\mu}$ is the stress tensor of $D^*_s$ and $V$ represents the vector meson field $J/\psi$ or $\phi$, $V^{\mu\nu}=\partial^{\mu}V^{\nu}-\partial^{\nu}V^{\mu}$~\cite{Ma:2010xx,Zhu:2021exs}.
Both $g_{J/\psi \bar{D}_{s0}D^{*}_{s}}=0.225\text{GeV}^{-1}$ and $g_{\phi\bar{D}_{s0}D^{*}_{s}}=0.135\text{GeV}^{-1}$ are coupling constants, and their numerical results are given in Ref.~\cite{Ma:2010xx}.

\begin{figure*}[htbp]
	\begin{center}
		\subfigure{}
		{
			\includegraphics[scale=0.65]{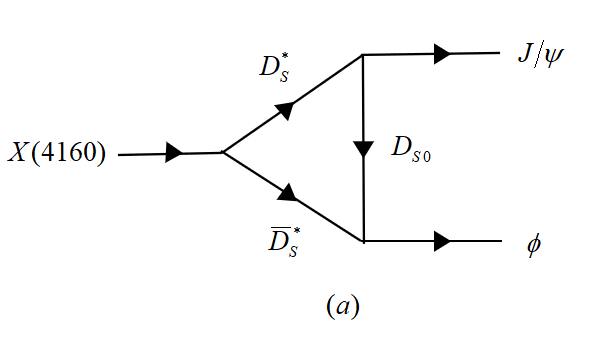}
		}
		\subfigure{}
		{
			\includegraphics[scale=0.65]{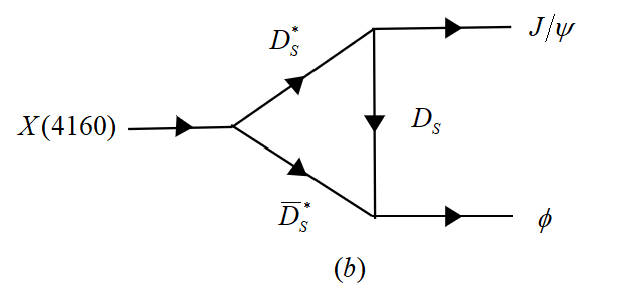}
		}
	\subfigure{}
	{
		\includegraphics[scale=0.65]{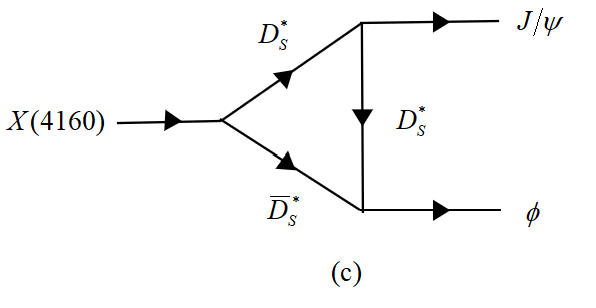}
	}
	\end{center}
	\vspace{-0.7cm}
	\caption{Feynman diagrams contributed to process $X(4160) \to J/\psi\phi$ by exchanging $D_{s0}$, $D_{s}$ and $D_{s}^*$ meson respectively.}
	\label{fig2}
\end{figure*}

For the $D_{s}D_{s}^{*}V$ vertices, the following effective Lagrangian get form Ref.~\cite{Zhu:2021exs,Oset:2002sh,Bramon:1992kr}
\begin{equation}
	\mathcal{L}_{VVP}=\frac{G}{\sqrt{2}} \epsilon^{\mu \nu \alpha \beta} \langle\partial_{\mu}V_{\nu}\partial_{\alpha}V_{\beta}P\rangle,
\end{equation}
where $G=3h^2/(4\pi^2f)$, $h=-G_{V}m_{\rho}/(\sqrt{2}f^2)$, $f=0.093$ GeV is the decay constant of $\pi$ meson, $G_{V}=0.069$ GeV, and $m_{\rho}$ is the mass of $\rho$ meson. $P$ and $V_{\mu}$ are the flavor SU(4) matrices of the pseudoscalar meson field and the vector meson field, respectively,
\begin{equation}
	\begin{aligned}
		P =\\ 
		\sqrt{2}&\left(\begin{array}{cccc}
			\frac{\pi^{0}}{\sqrt{2}}+\frac{\eta}{\sqrt{6}}+\frac{\eta^{\prime}}{\sqrt{3}} & \pi^{+} & K^{+} & \bar{D}^{0} \\
			\pi^{-} & -\frac{\pi^{0}}{\sqrt{2}}+\frac{\eta}{\sqrt{6}}+\frac{\eta^{\prime}}{\sqrt{3}} & K^{0} & -D^{-} \\
			K^{-} & \bar{K}^{0} & -\sqrt{\frac{2}{3}} \eta -\frac{1}{\sqrt{3}} \eta^{\prime} & D_{s}^{-} \\
			D^{0} & -D^{+} & D_{s}^{+} & \eta_{c}
		\end{array}\right)\\
		V_{\mu}&=\left(\begin{array}{cccc}
			\frac{1}{\sqrt{2}}\left(\rho^{0}+\omega\right) & \rho^{+} & K^{*+} & \bar{D}^{* 0} \\
			\rho^{-} & \frac{1}{\sqrt{2}}\left(-\rho^{0}+\omega\right) & K^{* 0} & -D^{*-} \\
			K^{*-} & \bar{K}^{* 0} & \phi & D_{s}^{*-} \\
			D^{* 0} & -D^{*+} & D_{s}^{*+} & J / \psi 
		\end{array}\right)_{\mu}.
	\end{aligned}
\end{equation}

In addition to the effective Lagrangian mentioned above, the following effective couplings between the charmonium/light mesons and the charmed mesons are also involved in our current estimates which can be constructed by the heavy quark limit and chiral symmetry \cite{Casalbuoni:1996pg,Colangelo:2003sa,Cheng:2004ru}:
	\begin{equation}
		\begin{aligned}
			\mathcal{L}_{D^{*} D^{*} J/{\psi}} =& i g_{D^{*} D^{*} J/{\psi}} {\psi}^{\mu}(D_{\mu i}^{* \dagger} \stackrel{\leftrightarrow}{\partial}_{\nu} D_{i}^{* \nu}+D_{\nu i}^{* \dagger} \stackrel{\leftrightarrow}{\partial}^{\nu} D_{\mu i}^{*}\\
			&-D_{i}^{* \dagger \nu} \stackrel{\leftrightarrow}{\partial}_{\mu} D_{\nu i}^{*}) \\
			\mathcal{L}_{D^{*} D^{*} V} =&i g_{D^{*} D^{*} V} V_{i j}^{\mu} D_{\nu i}^{* \dagger} \stackrel{\leftrightarrow}{\partial}_{\mu} D_{j}^{* \nu} \\
			& +4 i f_{D^{*} D^{*} V}\left(\partial^{\mu} V_{i j}^{\nu}-\partial^{\nu} V_{i j}^{\mu}\right) D_{\mu i}^{*} D_{j}^{* \dagger \nu}
		\end{aligned}
	\end{equation}
where $A \stackrel{\leftrightarrow}{\partial} B \equiv A \partial B-B \partial A$; $i,j$ are flavor indices; $V_{i j}$ represent the $3 \times 3$ vector meson fields in $SU(3)$ flavor symmetry. $D_{i}^{*}=\left(D^{*}_{0}, D^{* +}, D_{s}^{* +}\right)$is the triplet of vector $D^{*}$ mesons containing light antiquarks $\bar{u},s$ and $\bar{s}$, respectively. The chiral couplings $g_{D^*D^*J/\psi},g_{D^*D^*V}$ and $f_{D^*D^*V}$ are fixed as : $g_{D^*D^*V}=\beta g_{V}/\sqrt{2}, f_{D^*D^*V}=m_{D^*}\lambda g_{V}/\sqrt{2}, $ $g_{D^*D^*J/\psi}=(m_{D^*}m_{J/\psi})/(m_Df_{J/\psi})$, where $f_{J/\psi}=416.4$MeV is the $J/\psi$ leptonic decay constant; $g_{V} \approx 5.8$ and $\beta\approx 0.9$ are fixed using vector dominance; the parameter $\lambda=0.56\text{GeV}^{-1}$ is extracted by matching the heavy hadron chiral perturbation theory (HHChPT) to lattice QCD and light cone sum rules~\cite{Colangelo:2003sa}.

Based on the relevant effective lagrangians we have presented before, we can give the decay amplitudes of the Feynman diagrams in FIG.\ref{fig2}:
\begin{equation}
	\begin{aligned}
\mathcal{M}_{a}=&\int \frac{d^4q}{(2\pi)^4}(i)^3 [g_X\tilde{\Phi}_{[(\omega_{12}k_1-\omega_{21}k_2)^2]}\epsilon_{p}^{\mu\nu}]\\
&\times [G\epsilon^{\alpha\beta\rho\sigma}(-ik_1^{\alpha})(ip_1^{\rho})\epsilon_{p_1}^{\sigma}][G \epsilon^{\theta\tau\phi\zeta}(-ik_2^{\theta})(ip_2^{\phi})\epsilon_{p_2}^{\zeta}]\\
&\times \frac{-g^{\mu\beta}+\frac{k_1^{\mu}k_1^{\beta}}{m_1^2}}{k_1^2-m_1^2} \frac{-g^{\nu\tau}+\frac{k_2^{\nu} k_2^{\tau}}{m_2^2}}{k_2^2-m_2^2} \frac{1}{q^2 -m_q^2}
\\
		\mathcal{M}_{b} =&
		\int \frac{d^4q}{(2\pi)^4}(i)^3 [g_X\tilde{\Phi}_{[(\omega_{12}k_1-\omega_{21}k_2)^2]} \epsilon^{\mu\nu}_{p}]\\
		&\times [g_{J/\psi D_{s0}D^{*}_{s}}(-ik_1^{\alpha}g^{\beta\rho}+ik_1^{\beta}g^{\alpha\rho})(ip_1^{\alpha}g^{\beta\sigma}-ip_1^{\beta}g^{\alpha \sigma})\epsilon_{p_1}^{\sigma}]\\
		&\times [g_{\psi D_{s0}D^{*}_{s}}((-i)k_2^{\theta}g^{\tau \phi}+ik_2^{\tau}g^{\theta\phi})(ip_2^{\theta}g^{\tau \zeta}-ip_2^{\tau}g^{\theta \zeta})\epsilon_{p_2}^{\zeta}] \\
		&\times \frac{-g^{\mu\rho}+\frac{k_1^{\mu}k_1^{\rho}}{m_1^2}}{k_1^2-m_1^2} \frac{-g^{\nu\phi}+\frac{k_2^{\nu} k_2^{\phi}}{m_2^2}}{k_2^2-m_2^2} \frac{1}{q^2 -m_q^2}  \\
		\mathcal{M}_{c}=& \int \frac{d^4q}{(2\pi)^4}(i)^3[g_X \tilde{\Phi}_{[(\omega_{12}k_1-\omega_{21}k_2)^2]}\epsilon^{\mu\nu}_{p}] [g_{D^* D^* J/\psi }]\\
		&\times [\epsilon^{a}_{p_1}g^{\alpha a} (g^{\beta\rho}g^{\sigma\alpha}(q^{\beta}+k_1^{\beta}) + g^{\beta\sigma}g_{\alpha\rho}(q^{\beta}+k_1^{\beta})\\
		&-g^{\beta \rho}g^{\beta\sigma}(q^{\alpha}+k_1^{\alpha}))] [\epsilon^{x}_{p_2}g^{\theta x}g_{D^* D^* V}(q^{\theta}-k_2^{\theta})g^{\tau\xi}g^{\tau\phi}\\
		&+4f_{D^*D^*V}g^{\theta\phi}g^{\tau\xi}(p_2^{\tau}g^{\theta x}-p_2^{\theta}g^{\tau x})\epsilon_{p_2}^x] \\
		&\times 
		\frac{-g^{\mu\rho}+\frac{k_1^{\mu}k_1^{\rho}}{m_1^2}}{k_1^2-m_1^2} \frac{-g^{\nu\phi}+\frac{k_2^{\nu} k_2^{\phi}}{m_2^2}}{k_2^2-m_2^2} \frac{-g^{\sigma\xi}+\frac{q^{\sigma}q^{\xi}}{m_q^2}}{q^2 -m_q^2},
	\end{aligned}
\end{equation}
where the correlation function $\tilde{\Phi}$ takes the form of Eq.~(\ref{eq2}), $\mathcal{M}_{a}$ , $\mathcal{M}_{b}$ and $\mathcal{M}_{c}$ represent the decay amplitude of diagrams in FIG.~\ref{fig2} (a), (b) and (c), respectively.

We label the four-momentum of $X(4160)$, $D^*_s$, $\bar{D}^*_s$, $J/\psi$ and $\phi$ as $p, k_1, k_2, p_1$ and $p_2$, respectively.
$\epsilon_{p}^{\mu\nu}$ and $\epsilon^{\mu}$ are the polarization vectors for spin-$2$ and spin-$1$ mesons. We use the following sum of the polarization vectors at present work~\cite{LopezCastro:1997im} :
\begin{equation}\label{eq8}
	\begin{aligned}
		& \sum_{\text {polar }} \epsilon_{\mu_{1} \nu_{1}}(p) \epsilon_{\mu_{2} \nu_{2}}^{*}(p)\\
		=&\frac{1}{2}\left(\theta_{\mu_{1} \mu_{2}} \theta_{\nu_{1} \nu_{2}}+\theta_{\mu_{1} \nu_{2}} \theta_{\nu_{1} \mu_{2}}\right)
		-\frac{1}{3} \theta_{\mu_{1} \nu_{1}} \theta_{\mu_{2} \nu_{2}} , \\
		& \sum_{\text {polar }} \epsilon_{\mu}(p) \epsilon_{\nu}^{*}(p)\\
		=& -g_{\mu\nu}+\frac{p_{\mu}p_{\nu}}{m^2},
	\end{aligned}
\end{equation}
where $\theta_{\mu\nu}=-g_{\mu\nu}+\frac{p_{\mu}p_{\nu}}{m^2}$.

The total contribution to  $X(4160)\to J/\psi\phi$ process is:
\begin{equation}
	\mathcal{M}_{(X(4160)\to J/\psi\phi)}=\mathcal{M}_{a}+\mathcal{M}_{b}+\mathcal{M}_{c}.
\end{equation} 

\subsection{Decay of $X(4160)\to D^{*}\bar{D}^{*}, D^*\bar{D}$ and $D\bar{D}$}

\begin{figure}[h]
	\begin{center}
		\includegraphics[scale=0.65]{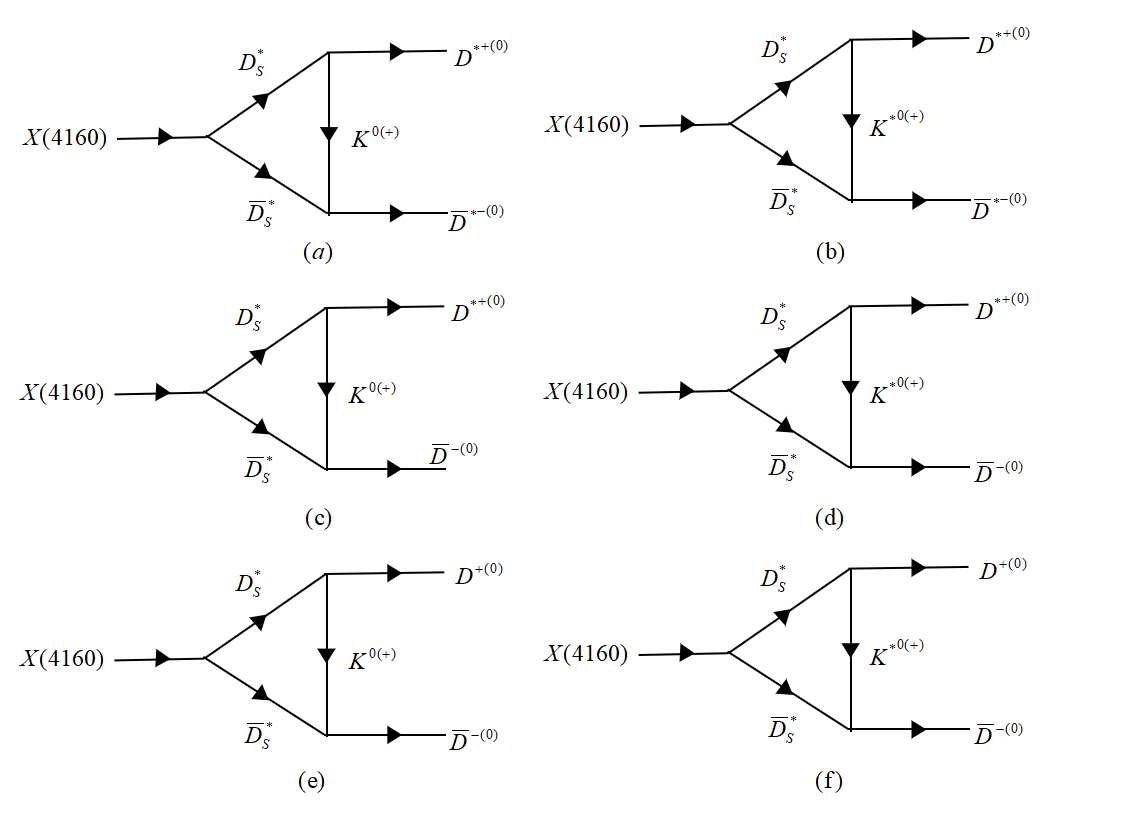}
		\vspace{-0.5cm}
		\caption{Feynman diagram contributed to process $X(4160) \to D^*\bar{D}^*/D^*\bar{D}/D\bar{D}$ by exchanging $K$ and $K^*$ mesons.}
		\label{fig3}
	\end{center}
\end{figure}

 In this section, we compute the decay process of $X(4160) \to D^{*}\bar{D}^{*}, D^*\bar{D}$ and $D\bar{D}$ and the relevant Feynman diagrams are shown in Fig.\ref{fig3}.

Based on the heavy quark limit and chiral symmetry \cite{Casalbuoni:1996pg,Colangelo:2003sa,Cheng:2004ru}, Lagrangians associate with light vectors and charm mesons can be constructed to describe the $D_s^*D^*K$ interaction, which are
\begin{equation}
	\begin{aligned}
		\mathcal{L}=&-i g_{D^{*} D P}\left(D_{i}^{\dag} \partial_{\mu} \mathcal{P}_{i j} D_{j}^{* \mu}-D_{i}^{* \mu \dag} \partial_{\mu} \mathcal{P}_{i j} D_{j}\right)\\
		&+\frac{1}{2} g_{D^*D^*P} \varepsilon_{\mu \nu \alpha \beta} D_{i}^{* \mu \dag} \partial^{\nu} \mathcal{P}_{i j} \stackrel{\leftrightarrow}{\partial^{\alpha}} D_{j}^{* \beta}\\
		&-i g_{D D V} D_{i}^{\dag} \stackrel{\leftrightarrow}{\partial}_{\mu} D^{j}\left(\mathcal{V}^{\mu}\right)_{j}^{i}\\
		&-2 f_{D^{*} D V} \epsilon_{\mu \nu \alpha \beta}\left(\partial^{\mu} \mathcal{V}^{\nu}\right)_{j}^{i}\left(D_{i}^{\dag} \stackrel{\leftrightarrow}{\partial}^{\alpha} D^{* \beta j}-D_{i}^{* \beta \dag} \stackrel{\leftrightarrow}{\partial^{\alpha}} D^{j}\right)\\
			&+i g_{D^{*} D^{*} V} D_{i}^{* \nu \dag} \stackrel{\leftrightarrow}{\partial}_{\mu} D_{\nu}^{* j}\left(\mathcal{V}^{\mu}\right)_{j}^{i}\\
		&+4 i f_{D^*D^*V} D_{i \mu}^{* \dagger}\left(\partial^{\mu} \mathcal{V}^{v}-\partial^{\nu} \mathcal{V}^{\mu}\right)_{j}^{i} D_{\nu}^{* j}+\text { H.c., }
	\end{aligned}
\end{equation}
where the $D^{(*)\dagger}=(\bar{D}^{(*)0} , D^{(*)-},D_{s}^{(*)-})$ is the triplets of pseudoscalar(vector)  charmed D meson, $\mathcal{P}$ and $\mathcal{V}_{\mu}$ are $SU(3)$ matrices.
In the heavy quark limit and chiral symmetry, the  charmed meson couplings  to the light vector and pseudoscalar mesons are,
\begin{equation}
\begin{aligned}
g_{D D V}=g_{D^{*} D^{*} V}=\frac{\beta g_{V}}{\sqrt{2}} \quad f_{D^{*} D V}=\frac{f_{D^{*} D^{*} V}}{m_{D^{*}}}=\frac{\lambda g_{V}}{\sqrt{2}}\\
g_{D^{*} D P}=\frac{2 g}{f_{\pi}} \sqrt{m_{D} m_{D^{*}}} \quad g_{D^{*} D^{*} P}= \frac{g_{D^{*} D \mathcal{P}}}{\sqrt{m_{D} m_{D^{*}}}},
\end{aligned}
\end{equation}
where the parameter $\beta =0.9$, $g_V=m_{\rho}/f_{\pi}$ with $f_{\pi}$ to be the decay constant of the $\pi$ meson \cite{Casalbuoni:1996pg,Chen:2019asm}, which value is presented in the previous section. The parameter $\lambda=0.56\text{GeV}^{-1}$ and $g=0.59$ can be obtained from the light cone sumrule and lattice QCD \cite{Isola:2003fh}.

The amplitude for the $X(4160) \to D^*\bar{D}^*$ in :

\begin{equation}
\begin{aligned}
\mathcal{M}_{a} &=(i)^3\int \frac{d^4q}{(2\pi)^4} \left[g_{X}\tilde{\Phi}_{[(\omega_{12}k_1-\omega_{21}k_2)^2]}\epsilon^{\mu\nu}_p \right] [\frac{1}{2}g_{D^*D^*P}\epsilon^{\alpha\beta\rho\sigma}\\
&(iq^{\beta})(-ik_1^{\rho}-ip_1^{\rho})]\\
&\times [\frac{1}{2}g_{D^*D^*P}\epsilon^{\theta\tau\phi\zeta}(-iq^{\tau})(ik_2^{\phi}+ip_2^{\phi})]\\
&\times \epsilon_{p1}^{\alpha}\epsilon_{p2}^{\zeta}
\frac{-g^{\mu\sigma}+\frac{k_1^{\mu}k_1^{\sigma}}{m_1^2}}{k_1^2-m_1^2} \frac{-g^{\nu\theta}+\frac{k_2^{\nu} k_2^{\theta}}{m_2^2}}{k_2^2-m_2^2} \frac{1}{q^2 -m_q^2} ,
\end{aligned}
\end{equation}

\begin{equation}
	\begin{aligned}
		\mathcal{M}_{b} = & (i)^3\int \frac{d^4q}{(2\pi)^4} \left[g_{X}\tilde{\Phi}_{[(\omega_{12}k_1-\omega_{21}k_2)^2]}\right]\\ &\times[g_{D^*D^*V}g^{\beta\rho}(-k_1^{\sigma}-p_1^{\sigma})
		+4f_{D^*D^*V}g^{\alpha\rho}\\
		&\times (q^{\alpha}g^{\beta\sigma}-q^{\beta}g^{\alpha\sigma})] [g_{D^*D^*V}g^{\tau\phi}(p_2^{\xi}+k_2^{\xi})\\
		&+4f_{D^*D^*V}g^{\theta\phi}(-q^{\theta}g^{\tau\xi}+q^{\tau}g^{\theta\xi})]\epsilon_{p}^{\mu\nu}\epsilon_{p_1}^{\rho}\epsilon_{p_2}^{\tau}\\
		&\times \frac{-g^{\mu\beta}+\frac{k_1^{\mu}k_1^{\beta}}{m_1^2}}{k_1^2-m_1^2} \frac{-g^{\nu\phi}+\frac{k_2^{\nu} k_2^{\phi}}{m_2^2}}{k_2^2-m_2^2} \frac{-g^{\sigma\xi}+\frac{q^{\sigma} q^{\xi}}{m_q^2}}{q^2 -m_q^2}.
\end{aligned}
\end{equation}

As for the process $X(4160) \to D^*\bar{D}$, we have:
\begin{equation}
	\begin{aligned}
		\mathcal{M}_{c} = & (i)^3\int \frac{d^4q}{(2\pi)^4} \left[g_{X}\Phi_{[(\omega_{12}k_1-\omega_{21}k_2)^2]} \right] \\
		 & \times \left[\frac{1}{2}g_{D^*D^*P}\epsilon^{\alpha\beta\rho\sigma}(iq^{\beta})(-ik_1^{\rho}-ip_1^{\rho})\right] \left[-ig_{D^*DP}(iq^{\theta})\right] \\
		 & \times 
		\frac{-g^{\mu\sigma}+\frac{k_1^{\mu}k_1^{\sigma}}{m_1^2}}{k_1^2-m_1^2} \frac{-g^{\nu\theta}+\frac{k_2^{\nu} k_2^{\theta}}{m_2^2}}{k_2^2-m_2^2} \frac{1}{q^2 -m_q^2}
		\epsilon_{p}^{\mu\nu}\epsilon_{p_1}^{\alpha},
	\end{aligned}
\end{equation}

\begin{equation}
	\begin{aligned}
		\mathcal{M}_{d} = & (i)^3\int \frac{d^4q}{(2\pi)^4} \left[g_{X}\Phi_{[(\omega_{12}k_1-\omega_{21}k_2)^2]} \right] \\
		& \times 
		[g_{D^*D^*V}g^{\beta\rho}(-k_1^{\sigma}-p_1^{\sigma})+4f_{D^*D^*V}g^{\alpha\rho} \\
		& \times (q^{\alpha}g^{\beta\sigma}-q^{\beta}g^{\alpha\sigma})] [-2f_{D^*DV}\epsilon^{\theta\tau\phi\xi}(q^{\theta})(p_2^{\phi}+k_2^{\phi})]\\
		& \times 	\frac{-g^{\mu\beta}+\frac{k_1^{\mu}k_1^{\beta}}{m_1^2}}{k_1^2-m_1^2} \frac{-g^{\nu\xi}+\frac{k_2^{\nu} k_2^{\xi}}{m_2^2}}{k_2^2-m_2^2} \frac{-g^{\sigma\tau}+\frac{q^{\sigma}q^{\tau}}{m_q^2}}{q^2 -m_q^2}
		\epsilon_{p}^{\mu\nu}\epsilon_{p_1}^{\rho}.
	\end{aligned}
\end{equation}

The amplitudes for the $X(4160) \to D\bar{D}$ corresponding to the diagrams is:
\begin{equation}
	\begin{aligned}
		\mathcal{M}_{e} =&(i)^3\int \frac{d^4q}{(2\pi)^4} \left[g_{X}\Phi_{[(\omega_{12}k_1-\omega_{21}k_2)^2]} \right] \\
		& \times  \left[-ig_{D^*DP}(iq^{\alpha})\right] \left[-ig_{D^*DP}(-iq^{\beta})\right]\\
		& \frac{-g^{\mu\alpha}+\frac{k_1^{\mu}k_1^{\alpha}}{m_1^2}}{k_1^2-m_1^2} \frac{-g^{\nu\beta}+\frac{k_2^{\nu} k_2^{\beta}}{m_2^2}}{k_2^2-m_2^2} \frac{1}{q^2 -m_q^2} \epsilon^{\mu\nu}_{p_1},
	\end{aligned}
\end{equation}
\begin{equation}
	\begin{aligned}
		\mathcal{M}_{f} = & (i)^3\int \frac{d^4q}{(2\pi)^4} \left[g_{X}\Phi_{[(\omega_{12}k_1-\omega_{21}k_2)^2]} \right] \\
		& \times \left[-2f_{D^*DV}\epsilon^{\alpha\beta\rho\sigma}(iq^{\alpha})(-ik_1^{\rho}-ip_1^{\rho})\right] \\
		& \times \left[-2f_{D^*DV}\epsilon^{\theta\tau\phi\xi}(-iq^{\theta})(-)(ip_2^{\phi}+ik_2^{\phi})\right]\\
		& \frac{-g^{\mu\sigma}+\frac{k_1^{\mu}k_1^{\sigma}}{m_1^2}}{k_1^2-m_1^2} \frac{-g^{\nu\xi}+\frac{k_2^{\nu} k_2^{\xi}}{m_2^2}}{k_2^2-m_2^2} \frac{-g^{\beta\tau}+\frac{q^{\beta}q^{\tau}}{m_q^2}}{q^2 -m_q^2}
		\epsilon_{p}^{\mu\nu}.
	\end{aligned}
\end{equation}

After considering the isospin doublet, the total contribution to  $X(4160)\to D^*\bar{D^*}$ process is:
\begin{equation}
\mathcal{M}_{(X(4160)\to D^*\bar{D^*})}=2(\mathcal{M}_{a}+\mathcal{M}_{b}).
\end{equation}

We can compute the decay width of $X(4160)\to J/\psi\phi / D^*\bar{D}^* / D^*\bar{D} / D\bar{D}$ separately by taking the decay amplitude into the following equation:
\begin{equation}
\Gamma_{X(4160)}= \frac{1}{2J+1} \frac{1}{8 \pi} \frac{|\vec{p}|}{M^2} \times |\mathcal{M}^{Tot}|^2 .
\end{equation}
where the $J$ and $M$ are the total angular momentum and mass of $X(4160)$ state, respectively. $\vec{p}$ is the three-momentum of the final state in the rest frame of the initial state.

\section{NUMERICAL RESULTS AND DISCUSSION}
\label{III}

The masses of all the particles involved in this work are presented in TABLE~\ref{table1}.
We study the strong decays of the $X(4160)$ to the two-body final states $J/\psi\phi$, $D^*\bar{D}^*$, $D^*\bar{D}$ and $D\bar{D}$ by assuming that it is a $D_s^*\bar{D^*_s}$ molecular state. For obtaining the two body decay widths by the triangle loop diagrams shown in FIG.~\ref{fig2} and FIG.~\ref{fig3}, we need to compute the coupling of $X(4160)$ to its constituents which can be estimated by the compositeness conditions given in Eq.~(\ref{eq5}). 
The value of the size parameter $\Lambda$ of hadron molecule is on the order of $1$ GeV 
and depends on different systems~\cite{Branz:2007xp,Faessler:2007us}, which is usually determined by experimental data.
Unfortunately, there are still too few experimental data available to determine  $\Lambda$. Therefore, we discuss the dependence of the coupling constants in the $\Lambda$ parameter range 0.8 to 1.2 GeV.

\begin{table}[h]
	\caption{The masses of the involved particles in units of GeV \cite{ParticleDataGroup:2014cgo}. }
	\label{table1}
	\begin{equation*}
		\begin{aligned}
			\begin{centering}
				\begin{array}{lc|lc|lc|lc}
					\hline \text{State} &\text{ Mass} &\text{State}&\text{Mass}&\text{State}&\text{Mass}&\text{State}&\text{Mass}\\
					\hline D^*_s& 2.112& D_s& 1.968&D_{s0}&2.317&J/\psi& 3.096\\
					\phi& 1.019& K^0&0.4976&K^{\pm}& 0.4936&D^{*0}&2.0069\\
					D^{*\pm}& 2.0102& D^{\pm} & 1.8695 &D^0 & 1.8648 &&\\
					\hline 
				\end{array}
			\end{centering}
		\end{aligned}
	\end{equation*}
\end{table}

\begin{figure}[h]
	\begin{center}
		\subfigure{}
		{
			\includegraphics[scale=0.8]{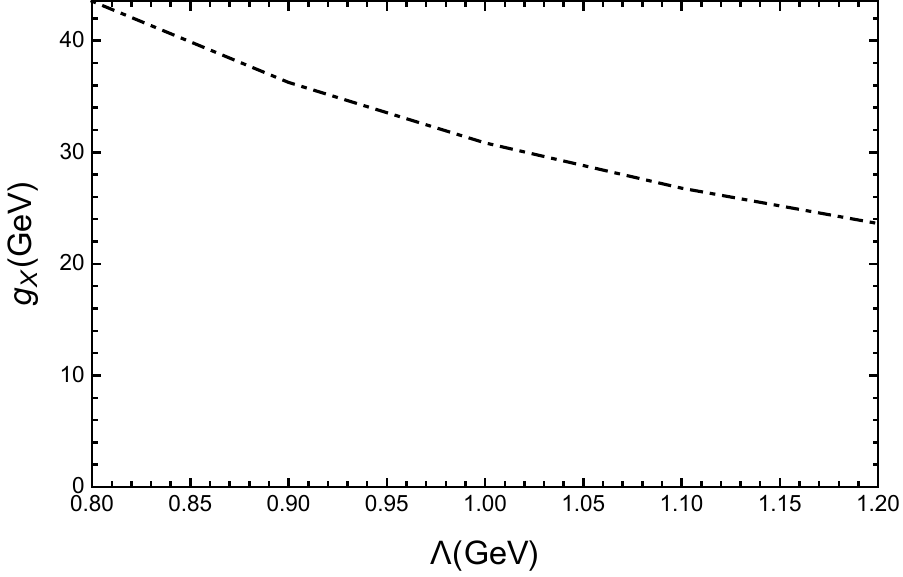}
		}
		\vspace{-0.5cm}
		\caption{The $\Lambda$ dependence of the coupling constant $g_X$. }
		\label{fig4}
	\end{center}
\end{figure}

\begin{figure}[h]
	\begin{center}
		\subfigure{}
		{
			\includegraphics[scale=0.8]{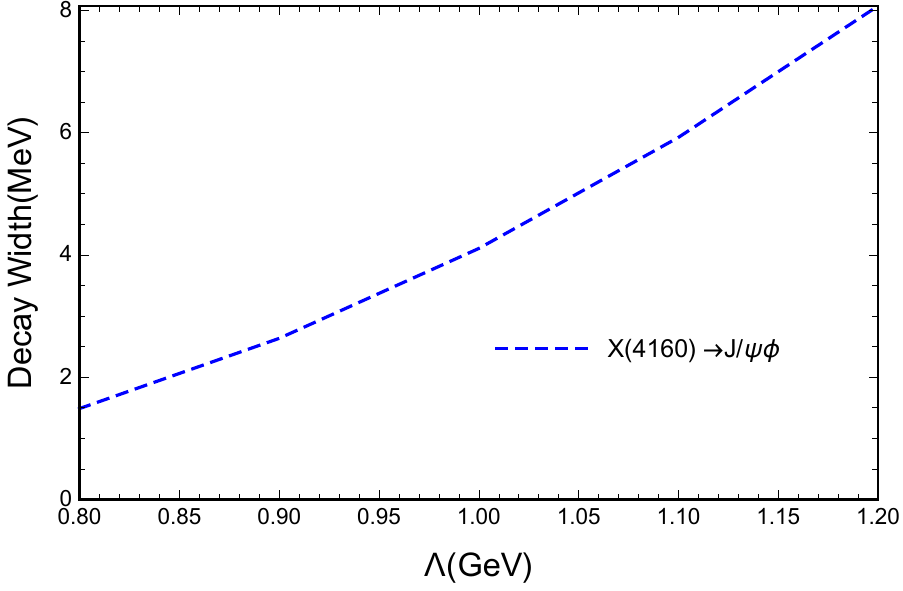}
		}
		\vspace{-0.5cm}
		\caption{The partical decay width of $J/\psi\phi$ decay modes  of $X(4160)$, which depends on the parameter $\Lambda$.}
		\label{fig5} \qquad
		\subfigure{}
		{
			\includegraphics[scale=0.8]{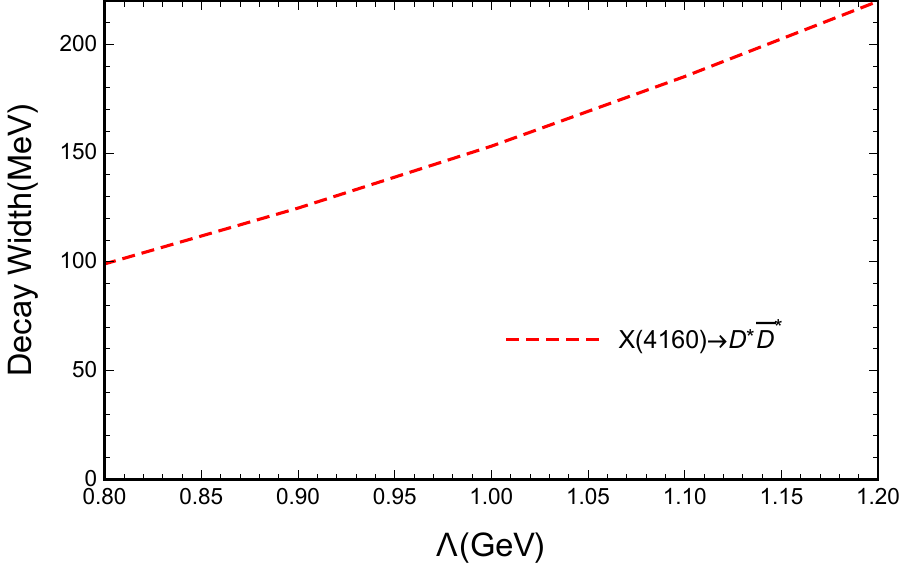}
		}
		\vspace{-0.5cm}
		\caption{The partical decay width of $D^*\bar{D}^*$ decay mode of $X(4160)$, which depends on the parameter $\Lambda$.}
		\label{fig6}
	\end{center}
\end{figure}
In FIG.~\ref{fig4}, we show the $\Lambda$ dependence of the couping constant, 
and we see $g_X$ decreases against the parameter $\Lambda$. The coupling constant between the $X(4160)$ particles and hadron molecules depend strongly on the value of the parameter $\Lambda$. In particular, when $\Lambda$ increases from $0.8$ to $1.2$ GeV, our numerical results roughly estimate that the coupling constant has a range of values:
\begin{equation}
	g_{X(4160) \to D^*_s\bar{D}^*_s} =  43.536 \sim 23.615.
\end{equation}

The decay width of the process $X(4160) \to J/\psi\phi$ has been showed in FIG.~\ref{fig5}. 
Our results show that the decay width of  $X(4160)\to J/\psi\phi$ is around $5$ MeV. 
Besides, this decay width depends on the parameter $\Lambda$, and increases with the increasing of $\Lambda$.
The decay width of the process $X(4160)\to D^*\bar{D}^*$ as a function of the size parameter $\Lambda$ is presented in FIG.~\ref{fig6}. 
The two-body decay in the $X(4160) \to D^*\bar{D}^*$ channel is also insensitive to the cut-off parameter $\Lambda$. The partial width of $X(4160)\to J/\psi\phi$ increases with the increasing $\Lambda$. 
The partial width of $X(4160)\to D^*\bar{D}^*$ also increases with the increasing of $\Lambda$. 
Within the parameter range under consideration, 
the partial widths for the $X(4160)$ to $J/\psi \phi$ and $D^*\bar{D^*}$ final states are $1.487 \sim 8.073 \text{MeV}$ and $98.993 \sim 219.734$ MeV, respectively.

Comparing the results in FIG.~\ref{fig5} and FIG.~\ref{fig6}, 
we find that transition $X(4160) \to D^*\bar{D}^*$ has a much larger decay branching ratio than $X(4160)\to J/\psi \phi$, because of the phase space and especially the couplings .
To observe such a state, the experimental observations for this transition of $X(4160) \to D^*\bar{D}^*$ may be essential. 

\red{}

\begin{figure}[h]
	\begin{center}
		\subfigure{}
		{
			\includegraphics[scale=0.65]{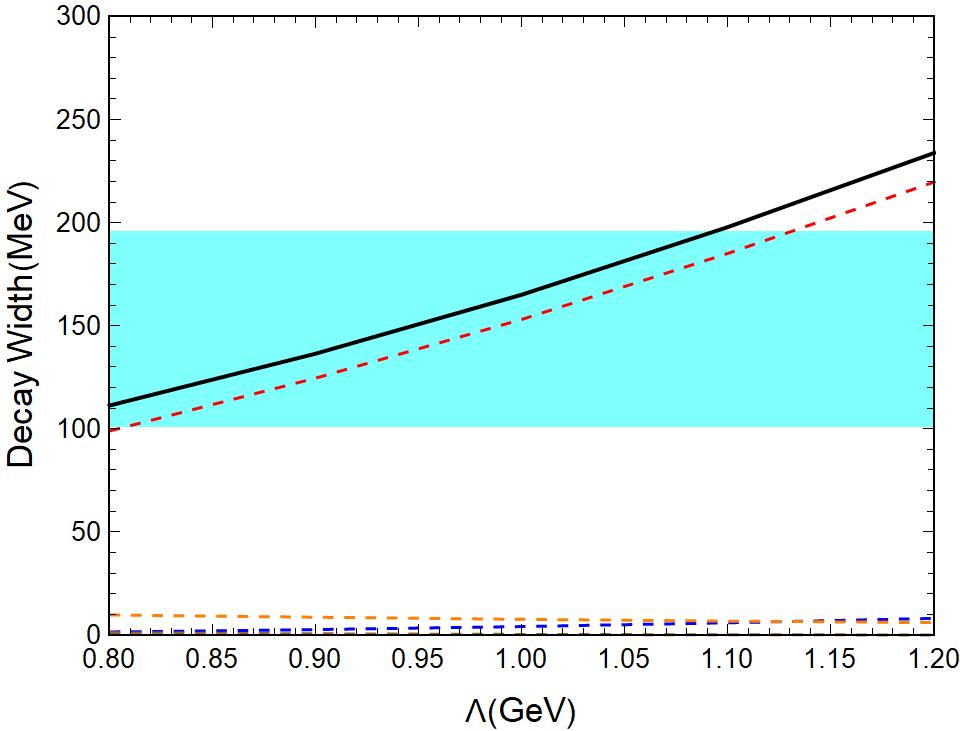}
		}
		\vspace{-0.5cm}
		\caption{ $X(4160) \to J/\psi\phi$[blue dashed], $D^*\bar{D}^*$[red dash-\newline ed], $D^*\bar{D}$[orange dashed], $D\bar{D}$[brown dashed] decays an-\newline d their sum[red soild] in dependence on $\Lambda$ and comparis-\newline on width data for $\Gamma_{X(4160)}$. The cyan band of "Exp" ar-\newline e the everage width of $X(4160)$ reported by LHCb Coll-\newline aboration.}
		\label{fig7}
	\end{center}
\end{figure}

\begin{figure}[h]
	\begin{center}
		\subfigure{}
		{
			\includegraphics[scale=0.65]{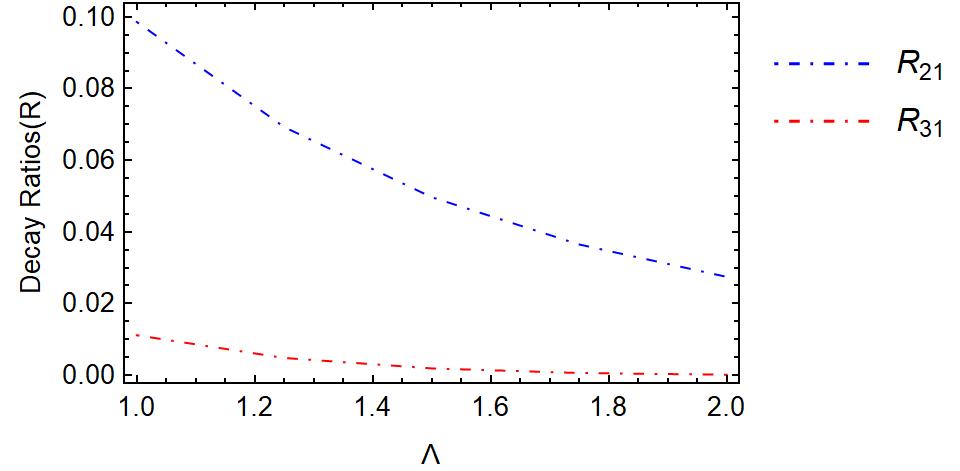}
		}
		\vspace{-0.5cm}
		\caption{ The $\Lambda$ dependences of the decay ratios.}
		\label{fig8}
	\end{center}
\end{figure}

 For a more visual presentation of the data, we plotted the dependence of the total decay width of $X(4160)$ and $\Lambda$ in FIG.~\ref{fig7}.
		In the range of $0.8\sim 1.2$ of $\Lambda$ parameter, 
		the total decay width is estimated to be $111.341 \sim 233.859$ MeV. 
		 From the numerical results in FIG.~\ref{fig7}, we can clearly see that the value of the theoretical total decay width of $X(4160)$ is in $\Lambda$ parameters between $0.8 \sim 1.2$ is in relatively good agreement with the experimental results, which could serves as an important test of the molecular picture. We can test if $X(4160)$ is a molecular state of $D_s^*\bar{D}^*_s$ by measuring more information about the decay widths of $X(4160)$ in subsequent experiments.
		 
We further considered the cases where the $J^P$ quantum numbers  of the $X(4160)$ meson is $X_1(1^+)$, and the coupling constant $g_{X_1}$ also depends on the estimated values of the parameter $\Lambda$. In the range of $0.8\sim 1.2$ of $\Lambda$, $g_{X_1}$ is estimated to be $5.742 \sim 4.622$. Additionally, the coupling constant $g_{X_1}$ decrease as $\Lambda$ increases. The partial width of process $X(4160)\to J/\psi\phi$ varies from 6.113 to 17.547 MeV with the variation of $\Lambda$ from 0.8 to 1.2 GeV. The partial width of the process $X(4160)\to D^*\bar{D^*}$ increases from 430.176 MeV within the considered $\Lambda$ parameter region. Based on the numerical results, the total width of the $X(4160)$ particle with quantum numbers $2^{++}$ offers a more accurate explanation of the experimental data.

We also compare the relative ratios of $R_{21}=\Gamma(X(4160) \to D^*\bar{D})/ \Gamma(X(4160) \to D^*\bar{D}^*)$ and  $R_{31} =\Gamma(X(4160) \to D\bar{D})/ \Gamma(X(4160) \to D^*\bar{D}^*)$, with the results shown in FIG.\ref{fig8}. Upper limits for the relative ratios $R_{21}$ and $R_{31}$ have been provided experimentally, which are $\Gamma(X(4160) \to D^*\bar{D})/ \Gamma(X(4160) \to D^*\bar{D}^*) < 0.22$ and $\Gamma(X(4160) \to D\bar{D})/ \Gamma(X(4160) \to D^*\bar{D}^*) < 0.09$, respectively. It can be clearly seen from FIG.\ref{fig8} that our results are consistent with the experiments~\cite{BaBar:2004oro}.

\section{summary}
\label{IV}

Recently, the decay properties of $X(4160)$ particle have been studied 
due to the experimental discovery of another $J/\psi \phi$ decay channel of $X(4160)$~\cite{LHCb:2021uow}. 
The assumption that $X(4160)$ could be explained as a $D^*_s\bar{D}^*_s$ molecular state
has been researched before in the $J/\psi\phi$ mass distribution of the $B^+ \to J/\psi\phi K^+$ reaction~\cite{Wang:2017mrt}.
In this paper, by using the effective Lagrangian approach, 
we calculate the partical decay widths for $X(4160)$ to $J/\psi\phi$, $D^*\bar{D}^*$, $D^*\bar{D}$ and $D\bar{D}$,
in which $X(4160)$ is assigned as a $D^*_s\bar{D}^*_s$ hadronic molecule. 
The coupling constant of $X(4160)$ to its components is fixed by the Weinberg compositeness condition 
and the decay width evaluations are performed in the parameter region considered to be:
\begin{equation}
	\begin{aligned}
		&\Gamma(X(41460)\to J/\psi\phi) = 1.487 \sim 8.073 \text{ (MeV) } , \\
		&\Gamma(X(4160) \to D^*\bar{D}^*) = 98.993 \sim 219.734 \text{ (MeV) }\\ 
		&\Gamma(X(4160) \to D^*\bar{D}) = 9.761 \sim 6.028 \text{ (MeV) }\\
		&\Gamma(X(4160) \to D\bar{D}) = 1.101 \sim 0.023 \text{ (MeV) }	.
	\end{aligned}
\end{equation}

  In summary, our calculation results predict the decay widths of $X(4160)$, 
and studying the decay behavior of these processes allows us to better explore the nature of $X(4160)$ particle.
We hope in the experiments, i.e., the LHCb, Belle and BESIII Collaborations can provide more information, to further determine the structure of $X(4160)$.

\begin{acknowledgments}
\noindent
The authors thank Maojun Yan for his valuable comments. XL is supported by the National Natural Science Foundation of China under Grant No. 12205002.
HS is supported by the National Natural Science Foundation of China (Grant No.12075043, No.12147205).
\end{acknowledgments}

\bibliography{ref}
\end{document}